\documentclass[12pt]{article}
\pdfoutput=1
\usepackage{graphicx,amssymb,amsfonts,amsthm,amsmath,amssymb,amscd,amstext, color}

\def\exp{\operatorname{exp}}
\def\tr{\operatorname{tr}}
\def\Vol{\operatorname{Vol}}

\begin{document}

\thispagestyle{empty}

\begin{flushright}
YITP-SB-15-20
\end{flushright}

\begin{center}
\vspace{1cm} { \LARGE {\bf
Thermal Corrections to R\'{e}nyi entropies for Free Fermions
}}
\vspace{1.1cm}

Christopher P.~Herzog and Michael Spillane \\
\vspace{0.8cm}
{ \it C.~N.~Yang Institute for Theoretical Physics \\
Department of Physics and Astronomy \\
Stony Brook University, Stony Brook, NY  11794}

\vspace{0.8cm}

\end{center}

\begin{abstract}
\noindent
We calculate thermal corrections to R\'{e}nyi entropies for free massless fermions on a sphere.  More specifically, 
we take a free fermion on $\mathbb{R}\times\mathbb{S}^{d-1}$ and calculate the leading thermal correction to the R\'{e}nyi entropies for a cap like region with opening angle $2\theta$.  By expanding the density matrix in a Boltzmann sum, the problem of finding the R\'{e}nyi entropies can be mapped to the problem of calculating a two point function on an $n$ sheeted cover of the sphere.  We follow previous work for conformal field theories to map the problem on the sphere to a conical region in Euclidean space.  By using the method of images, we calculate 
the two point function and recover the R\'{e}nyi entropies.  
\end{abstract}

\pagebreak
\setcounter{page}{1}

\newpage

\section{Introduction}

Entanglement entropy has become of interest to various communities in physics from condensed matter and quantum information \cite{Osborne,Vidal} to black holes and quantum gravity \cite{Bombelli,Srednicki}.  The discovery of a way of calculating entanglement entropy holographically produced interest in the AdS/CFT community as well \cite{Ryu}.  

In this paper, we use the conventional definition of entanglement entropy.  
We assume that the Hilbert space factors nicely with respect to two complementary spatial regions,
$A$ and $\bar A$.  
The reduced density matrix and R\'{e}nyi entropies are then defined as
\begin{align}
\rho_A &\equiv \tr_{\bar{A}}\rho, \\
S_{n} &\equiv \frac{1}{1-n}\log\, \tr(\rho_{A})^n.
\end{align}
The factor of $1/(1-n)$ in the the definition of the R\'{e}nyi entropy is convenient for taking a $n \to 1$ limit and recovering the entanglement entropy:
\begin{align}
S_{EE} \equiv -\tr \left[\rho_{A} \log(\rho_{A})\right] = \lim_{n\rightarrow 1}S_{n} \ .
\end{align}

It was argued in Ref.\ \cite{ChrisMichael} that for gapped systems at small temperature, thermal corrections to the
entanglement entropy  are  Boltzmann suppressed.  Further evidence in $d=1+1$ can be found in Refs.\ \cite{Azeyanagi,Barrella,Datta,Chen,Herzog2013}.  For general conformal field theories with temperature $1/\beta$ on a circle of perimeter $L$, the coefficient of the Boltzman factor was calculated \cite{CardyChris}:
\begin{align}
\delta S_n & \equiv  S_n(T)-S_n(0) 
%
= \frac{g}{1-n}\left[\frac{1}{n^{2\Delta-1}}\frac{\sin^{2\Delta}\left(\frac{\pi \ell}{L}\right)}{\sin^{2\Delta}\left(\frac{\pi \ell}{nL}\right)}-n\right]e^{-2\pi \beta \Delta/L}+o(e^{-2\pi \beta \Delta/L}),\\
\delta S_{EE} &\equiv S_{EE}(T)-S_{EE}(0) = 2 g \Delta\left[1-\frac{\pi \ell}{L}\cot\left(\frac{\pi \ell}{L}\right)\right]e^{-2\pi \beta \Delta/L}+o(e^{-2\pi \beta \Delta/L}),
\end{align}
where $g$ is the degeneracy of the first excited state, $\Delta$ is the smallest scaling dimension among the operators, and $\ell$ is the interval length.  (In order for these formulae to hold, the conformal field theory on the circle has to have a unique ground state and a mass gap.)

For higher dimensional conformal field theories on $\mathbb{S}^1 \times \mathbb{S}^{d-1}$, an analogous thermal correction to entanglement entropy is also known \cite{ChrisCFT}.  The result for the entanglement entropy for a cap on a sphere with polar angle $\theta$ and radius $R$ is given in general by the following integral:
\begin{align}
\label{delta}
\delta S_{EE} &= g\Delta I_d(\theta)e^{-\beta\Delta/R}+o\left(e^{-\beta\Delta/R}\right),\\
\label{Id}
I_d(\theta) &= 2\pi\frac{\Vol(S^{d-2})}{\Vol(S^{d-1})}\int_0^\theta d\theta' \frac{\cos(\theta')-\cos(\theta)}{\sin(\theta)}\sin^{d-2}(\theta') \ . 
\end{align}
The derivation of this result relies on a conformal transformation from the sphere to hyperbolic space.  
Unfortunately, there can be subtleties associated with boundary terms when this transformation is invoked.
In the case of conformally coupled scalars \cite{ChrisCFT, ChrisJun}, these boundary terms replace $I_d$ in the
result above with $I_{d-2}$.  The issue is that the conformal coupling requires a Gibbons-Hawking like term on the boundary.  The natural constant $\theta$ boundary is different from the boundary of hyperbolic space, and this difference contaminates the entanglement entropy. 
%
In this paper we study free fermions which have no such Gibbons-Hawking like term and consequently no subtleties associated with the boundary.  
Thus we expect and indeed find that the result (\ref{delta}) holds for massless free fermions.
As an added benefit, we also compute thermal corrections to R\'enyi entropies for fermions.

This paper is organized as follows.  First we briefly review the mapping used in Ref. \cite{ChrisJun} that maps from the multi-sheeted cover of the sphere to a wedge in flat space.  We then calculate the two point function using the method of images, from which we can read off the thermal corrections to R\'{e}nyi entropies.  Finally, we compare these calculations of entanglement and R\'{e}nyi entropies with numerical results for fermions in $d=2+1$ and $d=3+1$.

\section{R\'{e}nyi's for a General CFT}

A main result from Ref.\ \cite{ChrisJun} was a general equation for the thermal correction to the R\'{e}nyi entropy for a conformal field theory.  We assume that when the conformal field theory is placed on a ${\mathbb S}^{d-1} \times {\mathbb R}$, there is a unique ground state $|0 \rangle$ and a set of degenerate 
first excited states $|\psi_i\rangle$ with energy $E_\psi$.  
We divide the ${\mathbb S}^{d-1}$ into a spatial region $A$ and complement $\bar A$ and consider instead of ${\mathbb S}^{d-1} \times {\mathbb R}$, an $n$-sheeted branched cover of this spacetime where the branching is over the region $A$.  
The result from Ref.\ \cite{ChrisJun} is  
\begin{align}
\label{generalRenyi}
\delta S_n = \frac{n}{1-n}\sum_i\left(\frac{\langle\psi_i(z)\psi_i(z^\prime)\rangle_n}{\langle\psi_i(z)\psi_i(z^\prime)\rangle_1}-1\right)e^{-\beta E_\psi}+o(e^{-\beta E_\psi})
\end{align}
where $\psi_i$ is an operator that creates one of the first excited states.  
The point $z$ is in the far Euclidean future and the point $z'$ in the far Euclidean past.
The subscript $n$ indicates the two-point function is to be evaluated on this $n$-sheeted branched cover.
(Note that the result (\ref{generalRenyi}) could have been anticipated from a very similar result in 1+1 dimensional conformal field theories \cite{CardyChris}.)

In general, it is not clear how to evaluate $\langle \psi_i(z) \psi_i(z') \rangle_n$.  
However, if we restrict to the case where $A$ is a cap on a sphere of opening angle $2 \theta$, 
then we can take advantage of a conformal transformation that maps
the $n$-sheeted branched cover of ${\mathbb S}^{d-1} \times {\mathbb R}$ to $C_n \times {\mathbb R}^{d-2}$ where $C_n$ is an $n$-sheeted cover of the complex plane, branched over the negative real axis.  It is convenient to make the transformation in a couple of steps, as was outlined in Ref.\ \cite{ChrisJun}.
The first step takes the cap on ${\mathbb S}^{d-1}$ to a ball in ${\mathbb R}^{d-1}$ 
%
(see the appendix of  Ref. \cite{Candelas}):
\begin{align}
ds^2&=  -dt^2+dr^2+r^2d\Omega^2\\
&=\Omega^2\left(-d\tau^2+d\theta^2+\sin^2(\theta)d\Omega^2\right),
\end{align}
where
\begin{align}
t\pm r&= \tan\left(\frac{t\pm \theta}{2}\right),\\
\Omega&=\frac{1}{2}\sec\left(\frac{\tau+\theta}{2}\right)\sec\left(\frac{\tau-\theta}{2}\right),
\end{align}
and $d\Omega^2$ is the line element on $\mathbb{S}^{d-2}.$  If the cap has opening angle $2 \theta_0$, then the
ball has radius $r_0 = \tan(\theta_0/2)$.  A further special conformal transformation maps the ball to a half space:
\begin{align}
y^\mu &= \frac{x^\mu-b^\mu x^2}{1-2b\cdot x+b^2 x^2}, \\
ds^2 = dy^\mu dy^\nu \delta_{\mu\nu} &=   (1-2b\cdot x+b^2 x^2)^{-2}dx^\mu dx^\nu \delta_{\mu\nu}.
\end{align}
We let $x^0$ and $y^0$ correspond to Euclidean times, and take $b^1 = 1/r_0$ to be the only non-vanishing value of the vector $b$.
After further rescaling and rotations, the inserted operators can be placed at $y'=$ (1,2$\theta_0,\vec{0}$) and  $y=$ (1,0,$\vec{0}$), where we are using polar coordinates $(r,\theta)$ on the $C_n$.   
(For further details, see Ref.\ \cite{ChrisJun}.)

We will employ a method of images strategy for computing $\langle \psi_i(y) \psi_i(y') \rangle_n$  on $C_n \times {\mathbb R}^{d-2}$.  This strategy was already used successfully for the scalar in Refs.\ \cite{ChrisJun,Cardy}.  The idea is to compute the two-point function using the method of images on the orbifold ${\mathbb C} / {\mathbb Z}_m$ for general $m$ and then to obtain $\langle \psi_i(y) \psi_i(y') \rangle_n$ by analytic continuation, setting $n = 1/m$.  
As the method of images relies on the fact that the underlying equations of motion are linear, we do not expect this method will be useful for interacting field theories.

In the fermionic case, there are issues associated with nontrivial phases, 
signs and a choice of spin structure which we must address.
One issue, which we now review, is that rotations act nontrivially on spinor wave functions.  

%
%

\subsection{Rotation on Fermions}

For a Dirac fermion we know the effect of a rotation on the components of the spinor \cite{Peskin}.  A general Lorentz transformation in Euclidean signature, $\Lambda$, is given by
\begin{align}
&\psi(x)\rightarrow \Lambda_{1/2} \psi(\Lambda^{-1}x),\\
&\Lambda_{1/2} =\exp\left(\frac{1}{8}\omega_{\mu\nu}[\gamma^\mu,\gamma^\nu]\right), \\
&\{\gamma^\mu ,\gamma^\nu\} = 2\delta^{\mu\nu},
\end{align}
where $\omega_{\mu\nu}$ parameterizes the rotations and Lorentz boosts. For the case of interest we are only interested in rotations in (0,1) plane, for which the only non-vanishing components are $\omega_{01}=-\omega_{10}=\phi$.  The matrix exponential can be done simply and is given by
\begin{align}
\label{lambda}\Lambda_{1/2}(\phi) = \cos(\phi/2)+\sin(\phi/2)\gamma^0\gamma^1.
\end{align}
If we then define 
\begin{align}
\gamma^z = \gamma^0+i\gamma^1 \text{ and } \gamma^{\bar z} = \gamma^0-i \gamma^1,
\end{align}
then equation ($\ref{lambda}$) simplifies to
\begin{align}
\Lambda_{1/2}(\phi) = \frac{1}{2}\gamma^0(e^{-i\phi/2}\gamma^z+e^{i\phi/2}\gamma^{\bar z}).
\end{align}

A fact that we will rely on heavily moving forward is that $\gamma^0 \gamma^z$ and $\gamma^0 \gamma^{\bar z}$ are projectors:
\begin{align}
(\gamma^0 \gamma^z)^2 = 2 (\gamma^0 \gamma^z) \ , &\; \; \;
(\gamma^0 \gamma^{\bar z})^2 = 2 (\gamma^0 \gamma^{\bar z}) \ , \\ 
(\gamma^0 \gamma^z) (\gamma^0 \gamma^{\bar z}) = 0 \ , &\; \; \; (\gamma^0 \gamma^{\bar z})(\gamma^0 \gamma^z) = 0 \ . 
\end{align}

\section{Analytic calculation of R\'{e}nyi Entropies}

In flat space the fermion 2-point function is, up to normalization,
\begin{align}
\langle \bar \psi(y') \psi(y) \rangle &= \frac{\gamma^0\gamma^\mu(y-y')_\mu}{|y-y'|^d} \\
&= -\frac{1}{d-2}\gamma^0\gamma^\mu\frac{\partial}{\partial x^\mu} \frac{1}{|y-y'|^{d-2}} \ .
\end{align}

Following Ref.\ \cite{ChrisJun}, the Green's function on a wedge ${\mathbb C} / {\mathbb Z}_m \times {\mathbb R}^{d-2}$ can be calculated via the method of images.  The Green's function via the method of images is given by rotating one of the fermions by $2\pi k/m$ where $k$ indexes the wedges and $m$ is the number of wedges. In going between adjacent wedges an extra factor of (-1) is added due to the spin structure (for example see Ref.\ \cite{Epple}).  The result is then 
\begin{align}
\langle \bar \psi(y') \psi(y) \rangle_{1/m} = -\frac{\gamma^0\gamma^\mu\partial_\mu }{d-2}\sum_{k=0}^{m-1} \frac{(-1)^k\Lambda_{1/2}(2\pi k /m)}{[|z-e^{2\pi i k/m}z'|^2
+(\mathbf{y}-\mathbf{y}')^2]^{(d-2)/2}} \ .
\end{align}
(Curiously, this expression only makes sense for $m$ an odd integer.  Nevertheless, we find that knowing the two-point function for odd integers is in general sufficient to make the analytic continuation to $n = 1/m$.)  
In the case of interest $\mathbf{y} = \mathbf{y}' = 0$, $z'$ = $e^{2 i \theta}$ and $z\rightarrow 1$, the two-point function 
can be rewritten
\begin{align}
\hspace*{-1cm}G^F_{(1/m,d)}(2\theta) &=  -\frac{\gamma^0}{d-2}\lim_{z\rightarrow1}
\sum_{k=0}^{m-1} \left(e^{\frac{-\pi i k (m-1)}{m}}\gamma^z\partial_z+e^{\frac{\pi i k (m-1)}{m}}\gamma^{\bar z}\partial_{\bar z}\right)\frac{1}{|z-e^{2 i(\pi  k/m+\theta)}|^{d-2}} \\
\label{Gf}&= \frac{\gamma^0}{4(d-2)}\left[\gamma^z\left((d-2)-i\partial_\theta\right)\sum_{k=0}^{m-1} \frac{e^{\frac{-\pi i k (m-1)}{m}}}{|1-e^{2 i(\pi  k/m+\theta)}|^{d-2}}
+\gamma^{\bar z} c.c.\right] \ .
\end{align}
From this expression, we can deduce the following recursion relation for the two-point function:
\begin{align}
\label{recursion}G^F_{(1/m,d+2)}(2\theta) &=
\left((\partial_\theta^2+d(d-2))\gamma^0(\gamma^z+\gamma^{\bar{z}})+2i(\gamma^0\gamma^z-\gamma^0\gamma^{\bar z})\partial_\theta\right)
\frac{G^F_{(1/m,d)}(2\theta)}{8d(d-1)}.
\end{align}

To obtain the R\'{e}nyi entropy we make the replacement $n=1/m$ in the two point function and we use that $E_\psi = \frac{d-1}{2R}$ for free fermions on a sphere of radius $R$:
\begin{align}
\delta S_n(\theta) &= \frac{n}{1-n}\sum_i \left(\frac{\langle\psi_i\psi_i\rangle_n}{\langle\psi_i\psi_i\rangle_1}-1\right)e^{-(d-1)\beta/(2R)}+o\left(e^{-(d-1)\beta/(2R)}\right),\\
&=  \frac{n}{1-n}\tr(G^F_{(n,d)}(2\theta) G^F_{(1,d)}(2\theta)^{-1}-1)e^{-(d-1)\beta/(2R)}+o(e^{-(d-1)\beta/(2R)}).
\end{align}

\subsection{d=2}

In $d=1+1$ we can choose gamma matrices ($\gamma^0 = \sigma^3$ and $\gamma^1 = \sigma^1$) such that
\[ \gamma^0\gamma^z = \left( \begin{array}{cc}
2 & 0  \\
 0 &0 \end{array} \right) \text{ and } 
 \gamma^0\gamma^{\bar z} = \left( \begin{array}{cc}
 0&0\\
 0& 2 \end{array} \right).\]
It is worth noting that it is convenient  to have $\gamma^0\gamma^z$ diagonal, but is not necessary.  Then the 2-point function is given by
\begin{align}
\begin{split}
G^F_{(1/m,2)}(2\theta) &=  \frac{1}{2}\sum_{k=0}^{m-1}\gamma^0\left( \gamma^z\frac{\exp(\frac{-i k\pi(m-1)}{m})}{1-\exp\left(2i(k\pi/m+\theta)\right)}+\gamma^{\bar z}\frac{\exp(\frac{i k\pi(m-1)}{m})}{1-\exp\left(-2i(k\pi/m+\theta)\right)} \right)        \\
&= \gamma^0\left(\gamma^z \frac{m i}{4}e^{-i \theta}\csc(m\theta)-\gamma^{\bar z} \frac{m i}{4}e^{i \theta}\csc(m\theta)\right) \ .
\end{split}
\end{align}

We can then calculate the R\'{e}nyi entropies (and entanglement).
\begin{align}
\begin{split}
\label{deltaSntwo}
\delta S_{n}(\theta) &= \frac{n}{1-n}\tr(G^F_{(n,2)}(2\theta) G^F_{(1,2)}(2\theta)^{-1}-1)e^{-\beta/(2R)}+o(e^{-\beta/(2R)}) \\
&=  \frac{2}{1-n} \left(\sin(\theta)\csc(\theta/n)-n \right)e^{-\beta/(2R)}+o(e^{-\beta/(2R)}) \ ,
\end{split}\\
\delta S_{EE}&= 2(1 -  \theta \cot(\theta))e^{-\beta/(2R)}+o(e^{-\beta/(2R)}) \ .
\end{align}
These agree with the known results for 2d CFTs \cite{CardyChris} in general and for 2d fermions \cite{Azeyanagi,Herzog2013}
in particular.

\subsection{d=4}
In $d=3+1$ we choose gamma matrices 
\[ \gamma^0 = \left( \begin{array}{cc}
\sigma^2 & 0 \\
0 & -\sigma^2\\
 \end{array} \right) \text{ and } 
 \gamma^1 = \left( \begin{array}{cc}
 \sigma^1&0\\
 0& -\sigma^1 \end{array} \right).\]
In this case the 2-point function is given by
\begin{align}
G^F_{(1/m,4)}(2\theta) &= \frac{i m}{8}(1+3m^2+(m^2-1)\cos(2m\theta))\csc^3(m\theta)\gamma^0\left(\gamma^z e^{-i\theta}-\gamma^{\bar z} e^{i\theta}\right) \ .
\end{align}

Repeating the calculation in $d=2$ we get
\begin{align}
\begin{split}
\label{deltaSnfour}
\delta S_{n}(\theta) &= \frac{n}{1-n}\tr(G^F_{(n,4)}(2\theta) G^F_{(1,4)}(2\theta)^{-1}-1)e^{-3\beta/(2R)}+o(e^{-3\beta/(2R)}) \\
&=  \frac{4}{(1-n)n^2} \left( (3+n^2-(n^2-1)\cos(2\theta/n))\csc^3(\theta/n)\sin^3(\theta)-4n^3 \right)e^{-3\beta/(2R)}
 \\
& \; \; \; \; \;  +o(e^{-3\beta/(2R)})  \ , 
\end{split} \\
\begin{split}
\delta S_{EE}(\theta) &= \lim_{n\rightarrow1} \frac{n}{1-n}\tr(G^F_{(n,4)}(2\theta) G^F_{(1,4)}(2\theta)^{-1}-1)e^{-3\beta/(2R)}+o(e^{-3\beta/(2R)})  \\
&= 2(5+\cos(2 \theta)-6 \theta \cot(\theta))e^{-3\beta/(2R)}+o(e^{-3\beta/(2R)}) \ .
\end{split}
\end{align}
The second result correctly reproduces the entanglement entropy correction found for general conformal field theories on the sphere \cite{ChrisCFT}.  The result for $\delta S_n(\theta)$ is new.

\subsection{d=3}

In odd dimensions we can choose the same $\gamma^z$ as we would use in one smaller dimension.  Namely,
\[ \gamma^0\gamma^z = \left( \begin{array}{cc}
2 & 0  \\
 0 &0 \end{array} \right) \text{ and } 
 \gamma^0\gamma^{\bar z} = \left( \begin{array}{cc}
 0&0\\
 0& 2 \end{array} \right).\]
Following previous work \cite{ChrisJun, Cardy}, we may try to convert the denominator of the Green's function (\ref{Gf}) to an integral in order to perform the sum over $k$.  
In the case of the scalar, the resulting expression can be analytically continued to all $m$ and thus in particular 
to $n = 1/m$.  
However, in the case of the fermion, some extra phases appear to spoil the analytic continuation.
We are able to extract thermal corrections 
to entanglement entropy from an $n \to 1$ limit of the integral successfully.  Thermal corrections to R\'enyi entropies remain out of reach however.

The first step in converting the sum to an integral is an integral representation of the cosecant used successfully in the analogous calculation for the scalars \cite{Cardy,ChrisJun}:
\begin{align}
\int_0^\infty dx\frac{x^{\theta/\pi+k/m-1}}{1+x}=\pi \csc (\theta+k\pi/m) \ .
\end{align}
From this integral representation, it directly follows that
\begin{align}
\label{sumd3}\sum_{k=0}^{m-1} \frac{e^{-i\pi k(m-1)/m}}{\sin(\pi k/m+\theta)} &= \frac{1}{\pi}\int_0^\infty dx \frac{x^{\theta/\pi-1}}{1+x}\sum_{k=0}^{m-1}x^{k/m}e^{-i \pi k(m-1)/m}\\
\label{int}&= \frac{1}{\pi}\int_0^\infty dx \frac{x^{\theta/\pi-1}}{1+x}\frac{(1+e^{-i\pi m}x)}{1+e^{i\pi/m}x^{1/m}} \ .
\end{align}
Using the representation ($\ref{int}$) in the Green's function ($\ref{Gf}$) for $d=3$, we obtain
\begin{align}
G^F_{(1/m,3)} = \frac{1}{8}(1-i \partial_\theta)\gamma^0\gamma^z \frac{1}{\pi} \int_0^\infty dx \frac{x^{\theta/\pi-1}}{1+x}\frac{e^{-i\pi m}(e^{i m\pi}+x)}{1+e^{i\pi/m}x^{1/m}}+\gamma^0\gamma^{\bar z} cc \ .
\end{align}

To get the entanglement entropy, we expand around $m=1$:
\begin{align}
G^F_{(1/m,3)} &= (1-i \partial_\theta)\frac{\gamma^0\gamma^z}{\pi}\int_0^\infty dx \, \frac{x^{\frac{\theta}{\pi}}}{1+x} \left(  \frac{1}{x}-\frac{\log(x)}{1-x}(m-1)+O(m-1)^2\right)  + cc  \\
%
%
&=\gamma^0\gamma^z \left(i e^{-i\theta}\csc^2(\theta)-\frac{2e^{i\theta}\pi}{(1+e^{i\theta})^3}(m-1)+O(m-1)^2)\right)+ cc \ . 
\end{align}
%
The entanglement entropy correction is then constructed from a ratio of Green's functions
\begin{align}
\delta S_{EE}(\theta) &= \lim_{m\rightarrow1} \frac{1}{m-1}\tr(G^F_{(1/m,3)}(2\theta) G^F_{(1,3)}(2\theta)^{-1}-1) e^{-\beta/R}+o(e^{-\beta/R})\\
&= 4\pi \csc(\theta)\sin^4(\theta/2)e^{-\beta/R}+o(e^{-\beta/R}) \ .
\end{align}
This result matches the general case derived in Ref.\ \cite{Herzog2013}.  

While this integral representation gives the correct thermal corrections to the entanglement entropy, it appears to fail for the R\'{e}nyi entropies.  
We suspect a reason is that the integral representation grows too quickly as a function of complex $m$ to satisfy the assumptions of Carlson's Theorem.  In other words, there will not be a unique analytic continuation.
We can break the integral up into two pieces, one from $0<x<1$ and a second from $1 < x <\infty$, and then replace the two integrals with double sums:
\begin{align*}
\begin{split}
\sum_{k=0}^{m-1} \frac{e^{-i\pi k(m-1)/m}}{\sin(\pi k/m+\theta)} = &\sum_{p,q=0}^\infty (-1)^{p+q} m \left(\frac{e^{-i m \pi}e^{-i \pi(q+1)/m}}{\pi(1+mp+q)-m\theta}\right. +
 \frac{e^{-i \pi(q+1)/m}}{\pi(1+m+mp+q)-m\theta} \\
 & \left. \hspace{1in} +\frac{e^{i \pi q/m}}{\pi(mp+q)+m\theta}+\frac{e^{-i m \pi}e^{i \pi q/m}}{\pi(m+mp+q)+m\theta} \right) \ .
\end{split}
\end{align*}
In the case of the scalar, the phases in the numerator of this expression vanish, and the sum has better convergence properties.  Here instead, for $m = iy$ pure imaginary, the sum has the same kind of growth as $\sin( \pi m )$, which vanishes for all integer $m$.\footnote{See appendix A for an alternate integral representation of the sum.}


%
%

%

\subsection{Recursion relation for Entanglement entropy}

We would also like to show that our recursion relation (\ref{recursion}) is compatible with the recursion relation for the entanglement entropy found in  Ref.\ \cite{Herzog2013}.  We start by Taylor expanding the two-point function and relating it to the entanglement entropy\footnote{Here we are taking the case where $\gamma^0\gamma^z$ is diagonal so that the inverse is particularly simple.}
\begin{align}
\label{A}G_{n,d}(2\theta) = G_d(2\theta)&+\delta G_d(2\theta)(n-1)+\mathcal{O}(n-1)^2,\\
G_d(2\theta) &= \gamma^0(\gamma^ze^{-i\theta}-\gamma^{\bar{z}}e^{i\theta})\frac{i\csc^{d-1}(\theta)}{2^d},\\
\delta S_{EE}(\theta) &= g \frac{\delta G_d(2\theta)}{G_d(2\theta)}e^{-\beta E_\psi}+o(e^{-\beta E_\psi}).
\label{dSEE}
\end{align}

We will proceed by induction and assume equation (\ref{delta}) 
in $d$ dimensions.  Equation (\ref{Id}) has the following recursion relation 
\begin{align}
I_d(\theta)-I_{d-2}(\theta) = -2\pi \frac{\Vol(S^{d-2})}{\Vol(S^{d-1})}\frac{\sin^{d-2}(\theta)}{(d-1)(d-2)}.
\end{align}
Then using equation (\ref{delta}) and recalling that $\Delta = (d-1)/2$, 
\begin{align}
\delta G_d(2\theta) &= \frac{d-1}{2} I_d(\theta) G_d(2\theta)\\
&=\frac{d-1}{2}G_d(2\theta)\left( I_{d+2}+ 2\pi \frac{\Vol(S^{d})}{\Vol(S^{d+1})}\frac{\sin^{d}(\theta)}{(d+1)d}\right).
\end{align}
Acting on both sides with the operator in equation (\ref{recursion}) and simplifying  yields
\begin{align}
\delta G_{d+2}(2\theta) &=\frac{d+1}{2} I_{d+2}(\theta) G_{d+2}(2\theta).
\end{align}
We checked the entanglement entropy for both $d=1+1$ and $d=2+1$. Thus by induction the two recursion relations are in agreement for both even and odd dimensions.

\section{Numerical Check}

We are interested in numerically checking our results. As mentioned earlier a free fermion on a sphere does not suffer from the same boundary term ambiguities as the conformally coupled scalar \cite{ChrisCFT}.  The numerics for a free fermion should then directly give the general conformal field theory results.
Using the convention $\bar{\Psi} = \Psi^\dagger \gamma^0 $ the Hamiltonian and Lagrangian densities for a fermion in curved space are given by $\Gamma$
\begin{align}
\mathcal{L} &= \sqrt{-g}\bar{\Psi}(i \gamma^{\lambda} D_\lambda)\Psi, \\
\mathcal{H} &= \sqrt{-g}\bar{\Psi}(i \gamma^{j} D_j)\Psi,\\
\{\Psi_\alpha(x)&,\Psi^\dagger_\beta(x')\}\sqrt{-g} = i \delta_{\alpha\beta}\delta(x-x')
\end{align}
Where $D_\lambda$ is the covariant derivative on the manifold.  This can be written explicitly in terms of the vierbein ($e^\lambda_I$) and spin connection ($\omega_{\lambda IJ}$).\footnote{%
 We use capital Roman letters $I, J, K, \ldots$ for flat space-time indices, lower case Greek $\lambda, \mu, \nu, \ldots$ for curved space-time indices, lower case Greek $\alpha, \beta, \gamma, \ldots$ for spinor indices, and lower case Roman $i, j, k, \ldots$ for curved spatial indices.
 }
We have defined the curved space gamma matrices and covariant spinor derivative
\begin{align}
\gamma^{\mu} &= \gamma^I e_I^\mu \ , \\
D_\mu &= \partial_\mu+\frac{1}{8}\omega_{\mu IJ}[\gamma^I,\gamma^J] \ ,
\end{align}
where $\omega_{\mu IJ}$ is the spin connection.  Using the torsion free Maurer-Cartan equation,
%
$
d e^i +e^j\wedge{\omega_j}^i =0
$, 
we can extract the spin connection, the nonvanishing elements of which are
\begin{align}
{\omega^j}_i = \cos( \theta_i) \left(\prod_{k=i+1}^{j-1}  \sin \theta_k \right) d \theta_j
\end{align}
The general Hamiltonian in $d+1$ dimensions is then
\begin{align}
\mathcal{H} = i\sqrt{-g} \bar{\Psi}  \sum_{\ell=1}^d \gamma^\ell\left(\prod_{j=1}^{\ell-1}\csc(\theta_j)\right)\left (\partial_{\theta_\ell}+\frac{d-\ell}{2}\cot(\theta_\ell)\right)\Psi
\end{align}
where $\ell$ is a flat spatial index.

We can remove the cotangents and the volume factor in the commutation relation with the following definition $\Psi = \left(\prod_{j=1}^d \csc^{(d-j)/2}(\theta_j)\right)\psi$: 
\begin{align}
\{\psi_\alpha(x),\psi_\beta(x')\} &=i\delta_{\alpha\beta}\delta(x-x') \ ,\\
\mathcal{H} = \sum_{i=1}^d\prod_{j=1}^{i-1}\csc(\theta_j)\bar{\psi} \gamma^i\partial_{\theta_i}\psi
&\equiv \bar{\psi}\left(\gamma_1\partial_{\theta_1}+\frac{1}{\sin(\theta_1)}\mathcal{O}_d\right) \psi \ .
\end{align}
To obtain a numerical result efficiently, we turn this Hamiltonian density into a $d=1+1$ Hamiltonian. To this end, we integrate over $\theta_i$ for $i>1$. 
We then calculate the spectrum of $\mathcal{O}$ where the lowest energy, smallest eigenvalue, gives the lowest order thermal correction to the R\'{e}nyi entropies.  For general $d$ the result is
\begin{align}
\label{Hamiltonian} H_{d}&= \int_0^\pi d\theta_1 \psi^\dagger \left(\gamma_0\gamma_1\partial_{\theta_1}+\frac{(d-2)\gamma_0}{2\sin(\theta_1)}\right) \psi 
\end{align}
as can be found in Ref. \cite{Camporesi}.

\begin{figure}[!htb]
\centering
  \centering
  \includegraphics[width=0.48\linewidth]{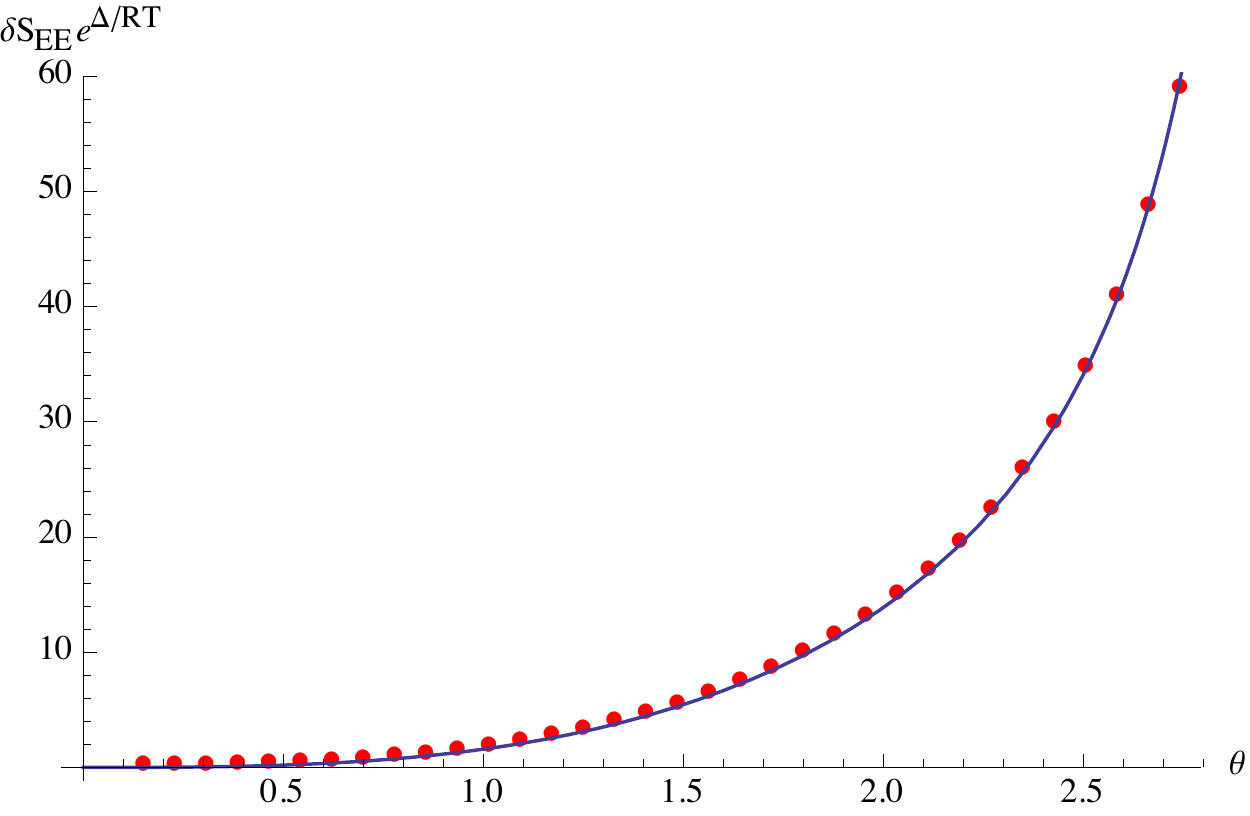}
  \quad
  \includegraphics[width=0.48\linewidth]{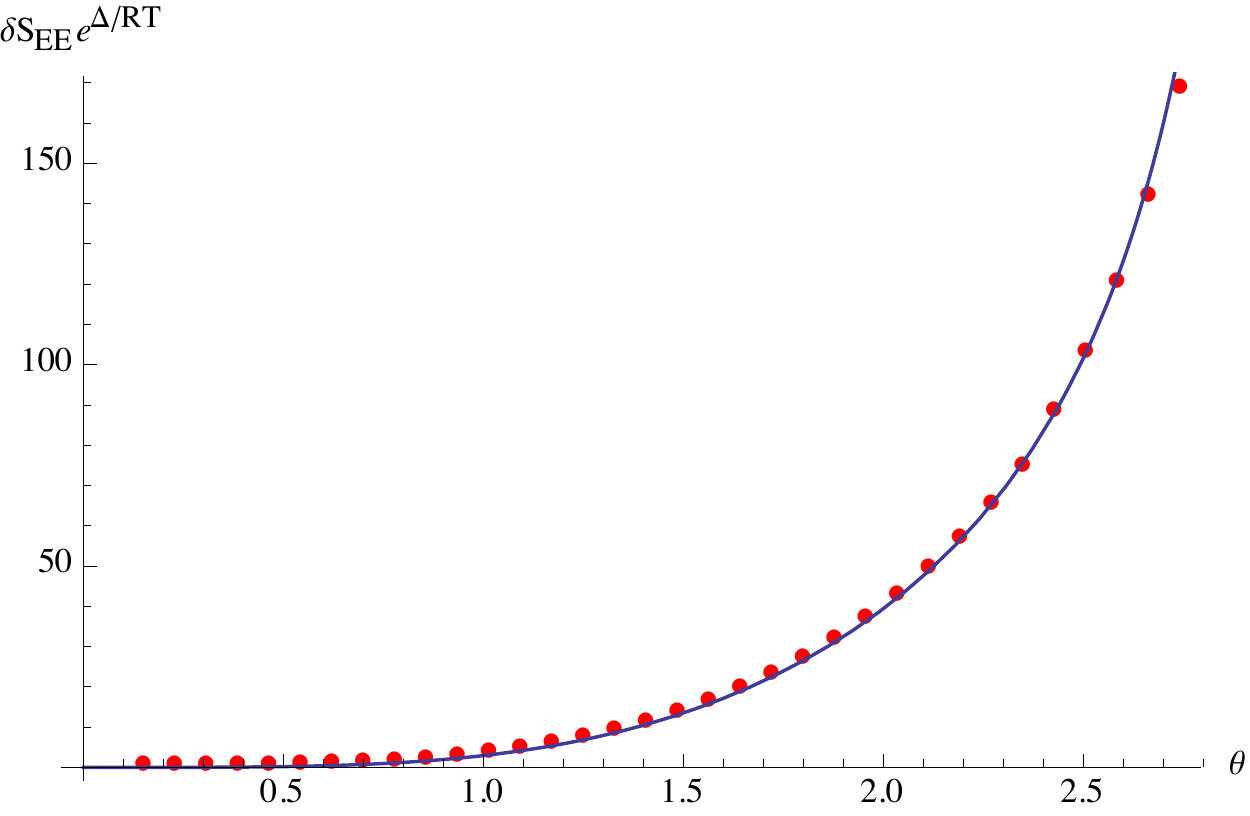}
 \caption{
 $\delta S_{EE}$ in $d=2+1$ (left), $3+1$  (right) with $200$ grid points.}
 \label{figEE}
\end{figure}

\begin{figure}[!htb]
\centering
  \centering
  \includegraphics[width=0.48\linewidth]{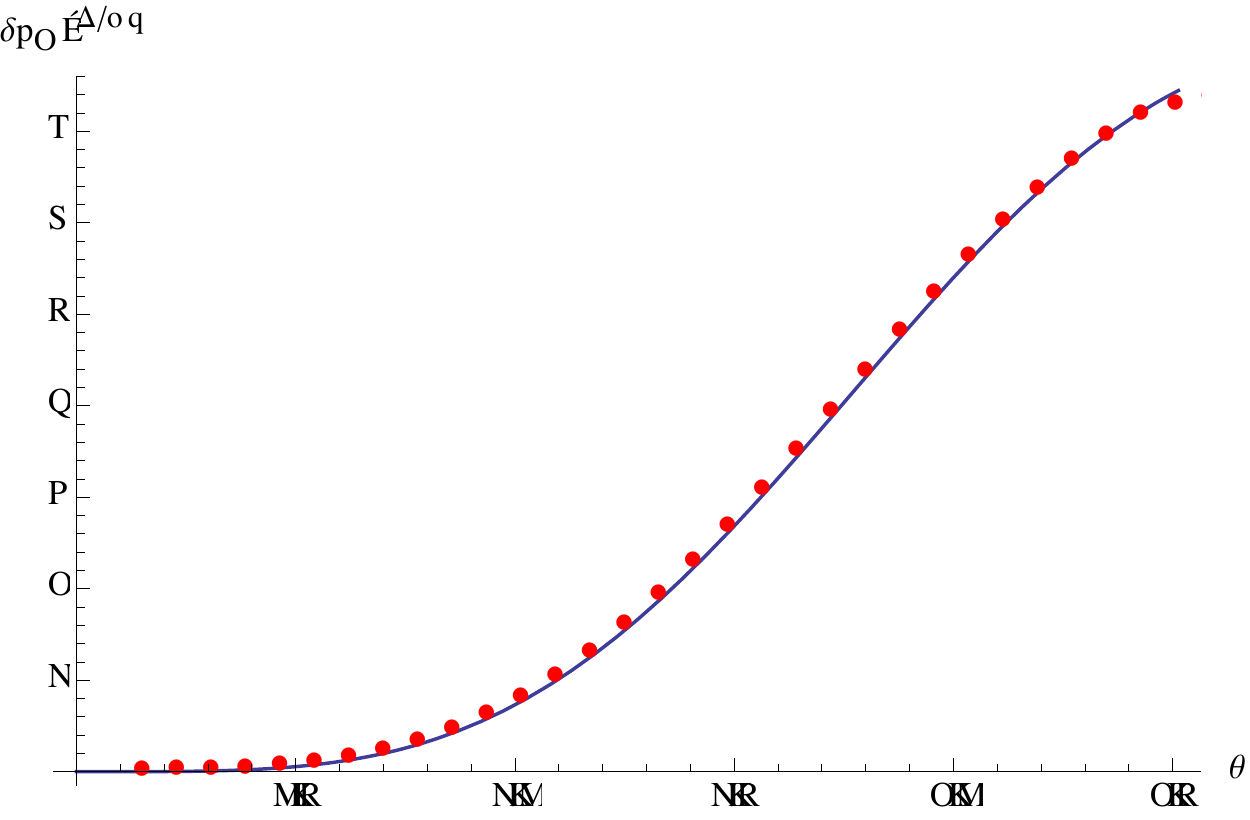}
  \quad
  \includegraphics[width=0.48\linewidth]{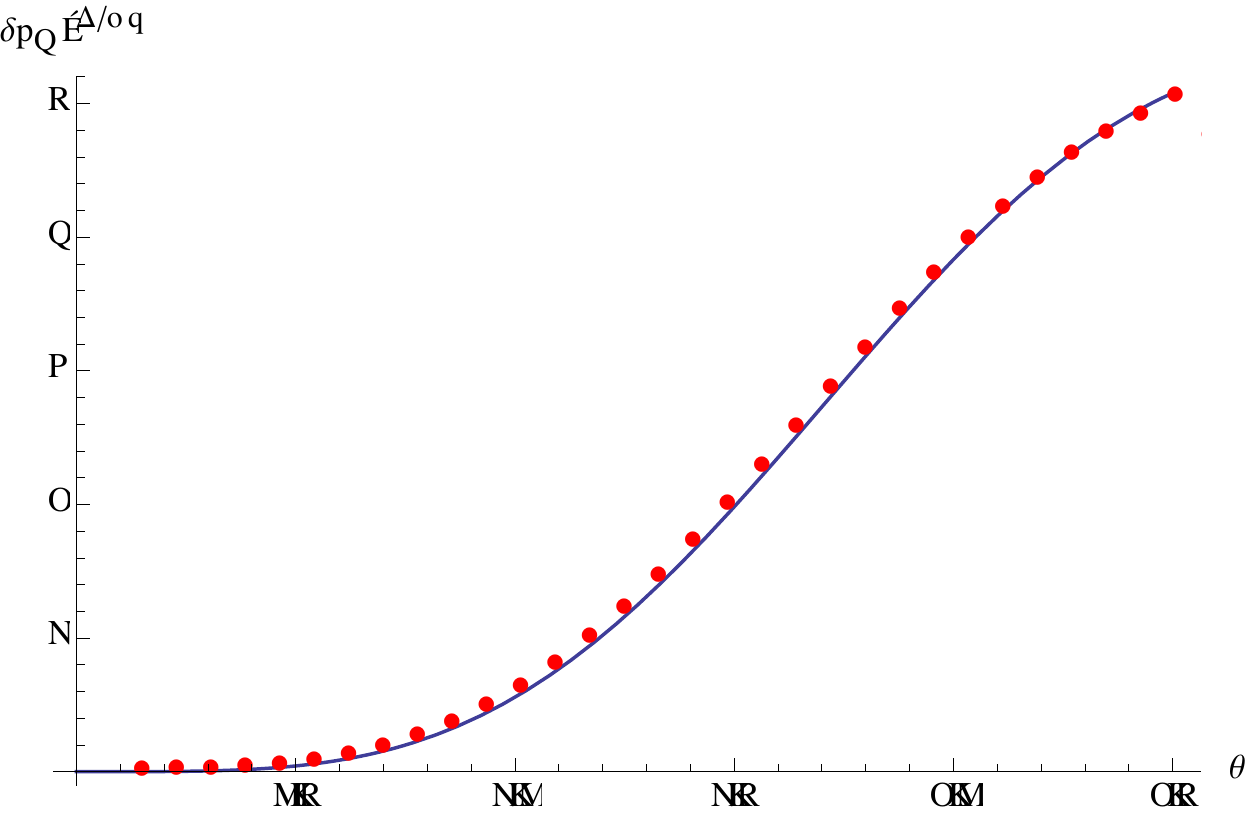}
 \caption{For $d=3+1$,
 $\delta S_2$ (left) and $\delta S_4$  (right) with $200$ grid points.}
 \label{figRenyi}
\end{figure}

We discretize these Hamiltonians, turning the integral into a sum and the derivative into a finite difference.  We then numerically calculate the entanglement and R\'{e}nyi entropies in the same way as previous papers \cite{ChrisJun,Herzog2013,EislerPeschel}.  We find agreement with our analytical results for the entanglement entropy in both $d=2+1$ and $d=3+1$ (see Figure~\ref{figEE}), and for the R\'{e}nyi entropy in $d=3+1$ (see Figure~\ref{figRenyi}).  

%

\section{Discussion}

In this paper we extended to include massless free fermions the work in Ref.\ \cite{ChrisJun}, which considered thermal corrections to R\'enyi and entanglement entropies for the conformally coupled scalar.  This extension allowed a direct and successful comparison with Ref.\ \cite{ChrisCFT} -- which provided general results for thermal corrections to entanglement entropy for conformal field theories -- without additional complications caused by boundary terms present for the conformally coupled scalar.   We also were able to calculate thermal corrections to the R\'{e}nyi entropies for the free fermion in even dimensions.  We give the analytic result in $d=1+1$ (\ref{deltaSntwo}) and $d=3+1$ (\ref{deltaSnfour}) along with a recursion relation (\ref{recursion}) which allows for computations of all even dimensions.  In odd dimensions, we were unable to find an analytic continuation that would allow us to calculate thermal corrections to R\'{e}nyi entropies, but we were able to reproduce the thermal corrections to entanglement entropy.  Amusingly, the situation is usually reversed, where one can compute R\'enyi entropies but the analytic continuation to entanglement entropy is not feasible.

It is possible that the methods used in this paper and those in Ref. \cite{ChrisJun} could allow for corrections to be calculated for other free higher spin theories and possibly more generally for conformal field theories.

\section*{Acknowledgments}
We would like to thank Jun Nian and Kristan Jensen for useful discussions.  We thank Ricardo Vaz for collaboration during the early stages of this project.  We also thank Martin Ro\v{c}ek for help in finding the sum in Appendix A.
C.~H. and M.~S. were supported in part by the National Science Foundation under Grant No.\ PHY13-16617.  

\appendix
\section{Alternate Formulation of d=3 Sum}

In our effort to find the R\'{e}nyi entropies in odd dimensions we came across an alternate form of the sum in Equation (\ref{sumd3}).  
\begin{align}
\label{finalsum}\sum_{k=0}^{m-1} \frac{e^{-\pi i k(m-1)/m}}{\sin(\pi k/m+\theta)}=  e^{-i\theta}\left(\cot(m\theta)+i-2 \frac{1}{\sin(m\theta)}\sum_{k=1/2}^{m/2-1}\frac{\sin[2k(\theta-\pi/2m)]}{\sin(\pi k/m)}\right).
\end{align}
This alternate representation is essentially a Fourier series on the shifted interval $\frac{\pi}{2m} < \theta < 2 \pi + \frac{\pi}{2m}$.

\begin{figure}[!htb]
\centering
  \centering
  \includegraphics[width=0.48\linewidth]{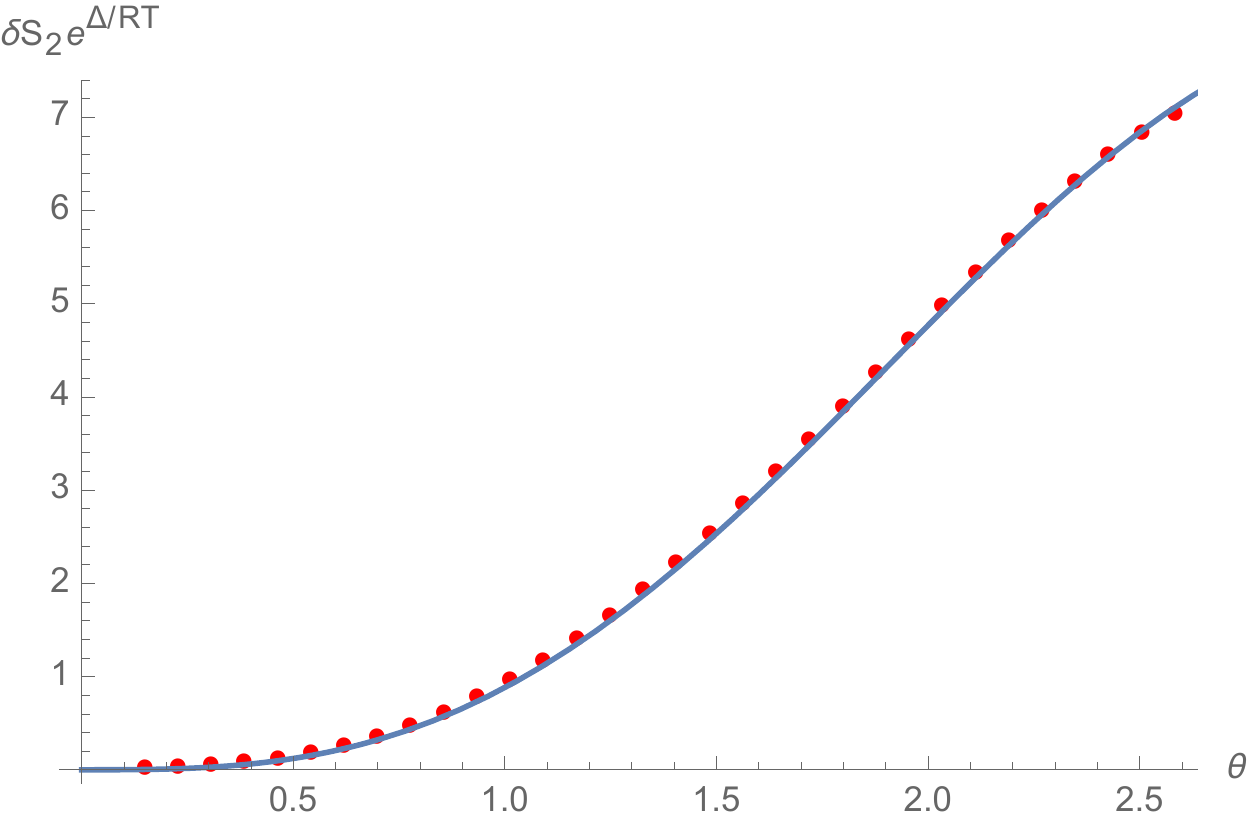}
 \caption{For $d=2+1$,
  $\delta S_2$ with $200$ grid points.}
 \label{Renyi3}
\end{figure}

Using the same integral form for the cosecant used for $d=3$ we can rewrite this new sum as an integral.
\begin{align}
\sum_{k=1/2}^{m/2-1}\frac{\sin[2k(\theta-\pi/2m)]}{\sin(\pi k/m)} = \frac{1}{\pi}\sum_{k=1/2}^{m/2-1}\int_0^\infty dx\frac{x^{k/m}}{x(1+x)}\sin[2k(\theta-\pi/2m)].
\end{align}
The integral can be evaluated for individual values of $n=1/m$.
This rewriting of the sum seems to have different issues with analytic continuation than those that plagued Equation (\ref{int}). It appears to reproduce correctly the $n=2$ thermal correction to the R\'{e}nyi entropy (see Figure~\ref{Renyi3}), but fails for the others and for the entanglement entropy.

Equation (\ref{finalsum}) gives the following results for the first couple of R\'{e}nyi entropies
\begin{displaymath}
\begin{array}{c|c}
  n=2 &  2 \sin(\theta/2)^3\\
  \hline
  n=3 & \frac{4}{3} \left[2 + \textrm{cos} \left(\frac{2 \theta}{3} \right) \right] \, \textrm{sin}^2 \left(\frac{\theta}{3}\right)
\end{array}
\end{displaymath}
(Note that the $n=3$ result is not reproduced by our numerics.)

\end{document}